# Multi-view SA-LA Net: A framework for simultaneous segmentation of RV on multi-view cardiac MR Images

Sana Jabbar[1], Syed Talha Bukhari[1], and Hassan Mohy-ud-Din, PhD[1,*]

[1]Department of Electrical Engineering, Syed Babar Ali School of Science and Engineering, LUMS, 54792, Lahore, Pakistan

*hassan.mohyuddin@lums.edu.pk

**Abstract.** We proposed a multi-view SA-LA model for simultaneous segmentation of RV on the short-axis (SA) and long-axis (LA) cardiac MR images. The multi-view SA-LA model is a multi-encoder, multi-decoder U-Net architecture based on the U-Net model. One encoder-decoder pair segments the RV on SA images and the other pair on LA images. Multi-view SA-LA model assembles an extremely rich set of synergistic features, at the root of the encoder branch, by combining feature maps learned from matched SA and LA cardiac MR images. Segmentation performance is further enhanced by: (1) incorporating spatial context of LV as a prior and (2) performing deep supervision in the last three layers of the decoder branch. Multi-view SA-LA model was extensively evaluated on the MICCAI 2021 Multi- Disease, Multi-View, and Multi- Centre RV Segmentation Challenge dataset (M&Ms-2021). M&Ms-2021 dataset consists of multi-phase, multi-view cardiac MR images of 360 subjects acquired at four clinical centers with three different vendors. On the challenge cohort (160 subjects), the proposed multi-view SA-LA model achieved a Dice Score of 91% and Hausdorff distance of 11.2 mm on short-axis images and a Dice Score of 89.6% and Hausdorff distance of 8.1 mm on long-axis images. Moreover, multi-view SA-LA model exhibited strong generalization to unseen RV related pathologies including Dilated Right Ventricle (DSC: SA 91.41%, LA 89.63%) and Tricuspidal Regurgitation (DSC: SA 91.40%, LA 90.40%) with low variance ($\sigma_{DSC}$: SA < 5%, LA< 6%).

## 1    Introduction

A report from World Health Organization (WHO) [1] predicts that, by 2023, approximately 23 million people will die due to cardiovascular diseases (CVDs). This will be a staggering increase of ~24% since 2019 which reported 18.6 million deaths worldwide due to CVDs [2]. These alarming statistics have sparked extensive research in the detection, staging, and treatment planning strategies of CVDs.

Cardiac MRI is the preferred imaging modality for diagnosis of CVDs. It provides high quality volumetric images of the heart noninvasively and without employing ionizing radiation. A routine cardiac MRI scan provides 3D scans of the end-diastolic (ED) and end-systolic (ES) cardiac phases. Clinical parameters of interest, including ES volume, ED volume, ejection fraction, and myocardial mass, are estimated from multi-phase cardiac MRI scans (i.e., 3D ED and ES images) and greatly aid in the diagnosis and prognostication of CVDs [3].

A pivotal step towards quantification of clinical parameters of interest is segmentation of the LV blood pool, RV blood pool, and LV Myocardium on multiphase cardiac MRI scans. The gold standard approach of multi-region segmentation of 3D cardiac MRI scans is manual segmentation by expert cardiologist. Manual segmentation is an extremely cumbersome process with high inter-observer and intra-observer variability [4]. To overcome limitations of manual segmentation, it is essential to propose fully automatic (or semi-automatic), robust, and accurate algorithms for multiregional segmentation of 3D cardiac MRI scans.

In the past decade, numerous cardiac image segmentation challenges have been launched from the platform of MICCAI including LVQuan18 [5], LVQuan19 [6], EMIDEC [7], and ACDC [8]. Most of these challenges sought fully automatic or semi-automatic methods that provide robust segmentation of multiple regions on cardiac MRI scans. Two important things are noteworthy. Firstly, Convolutional Neural Networks (CNNs) have demonstrated state-of-the-art performance in multiregional segmentation of 3D multi-phase cardiac MRI scans. Secondly, the principal focus of these challenges was accurate segmentation of the epicardial and endocardial boundaries (i.e., segmentation of the LV blood pool and LV myocardium) on cardiac MRI scans. LV segmentation has been the prime focus because noninvasive quantification of LV function is essential in the diagnosis of myocardial infarction, hypertrophy, cardiomyopathy, and dilated cardiomyopathy [9].

The most recent challenge hosted from MICCAI platform was the Multi-Centre, Multi-Vendor, and Multi-Disease Cardiac Image Segmentation Challenge (M&Ms-2020). M&Ms-2020 provided a large cohort of 375 subjects representing 6 different centers, 4 different vendors, and multiple cardiac pathologies [10]. The main aim of M&Ms-2020 was to come up with a fully automatic (or semi-automatic) segmentation method that generalizes well to novel datasets acquired at multiple centers, with multiple vendors, and representing diverse cardiac pathologies. M&Ms-2020 challenge concluded that deep learning solutions yielded state-of-the-art performance on cardiac MRI segmentation task with U-Net based architectures occupying top positions [11,12].

In 2021, a new challenge was hosted from the platform of MICCAI namely Multi-Disease, Multi-View, and Multi-Center Right Ventricular Segmentation in Cardiac MRI scans (M&Ms-2021) [13]. M&Ms-2021 specifically focused on the segmentation of RV blood pool which is essential to the study of RV related pathologies such as Dilated Right Ventricle, Tricuspid Insufficiency, Arrhythmogenesis, Tetralogy of Fallot and Interatrial Communication. Compared to LV, segmentation of RV blood pool is more challenging due to its highly complex and variable shape and ill-defined region boundaries on cardiac MRI scans. A novel aspect of M&Ms-2021 challenge is that, for each subject, it provides multi-view scans of the heart in the form of 3D short-axis images (SA) and 2D long-axis images (LA). 2D LA cardiac scans facilitate automatic definition of the basal plane of the RV which is often confused with the right atrium.

In this study, we proposed a multi-view deep neural network architecture for simultaneous segmentation of RV blood pool on SA and LA cardiac scans. The proposed architecture is called *multi-view SA-LA model*. The SA-LA model is a novel multi-encoder, multi-decoder architecture based on the U-Net model. One encoder-decoder pair segments the RV blood on SA images and the other pair on LA images. The main hypothesis of the study is that an extremely rich set of synergistic features can be assembled at the root of the encoder branch that combines feature maps extracted from matched SA and LA cardiac images. With extensive experiments on M&Ms-2021 dataset, we demonstrated that the proposed SA-LA model achieved state-of-the-art performance in the segmentation of RV blood pool on SA and LA cardiac images. Moreover, SA-LA model generalized to novel datasets, acquired at multiple centers with multiple vendors, and novel RV pathologies.

## 2 Material and Method

### 2.1 Cardiac MRI Dataset

The proposed SA-LA model was extensively evaluated on a publicly available dataset, namely M&Ms-2021 dataset, which forms the RV segmentation challenge in MICCAI 2021. M&Ms-2021 is a multi-disease, multi-view, and multi-center dataset for RV segmentation on cardiac MRI scans. M&Ms-2021 dataset is composed of 360 subjects acquired at 4 centers, with 3 vendors, and represents diverse RV related pathologies. M&Ms-2021 dataset is split into training (160 subjects), validation (40 subjects), and challenge (160 subjects) cohorts. The training and validation cohorts were publicly released and included 3D MRI scans of the ED and ES cardiac phases in the short-axis view and 2D MRI scans of the ED and ES cardiac phases in the long-axis view. Moreover, for the training cohort, manual segmentation of the LV blood pool (label=1), LV myocardium (label=2), and RV blood pool (label=3) on the short-axis and long-axis views were also provided. The challenge cohort was not publicly released by the organizers. To study the generalizability of the proposed approach, the validation and challenge cohorts included novel pathologies which were not present in the training cohort. The details of M&Ms-2021 dataset is summarized in **Table 1**.

## 2.2 Data Preprocessing

The 3D multiphase cardiac MRI scans were resampled to an in-plane resolution of $1.25 \times 1.25 \text{ mm}^2$ and cropped to an in-plane matrix dimension of $256 \times 256$ voxels. Furthermore, each 3D MRI volume was normalized to zero mean and unit-variance. For each subject, the 2D LA cardiac image slice (and the associated 2D manual segmentation image) was replicated to create a 3D volume of the same dimension as the corresponding 3D SA cardiac scan. This simple modification helped create a (subject-wise) matched SA-LA dataset of the same dimension which was subsequently used for training a 2D (slice-based) SA-LA model.

For accurate segmentation of RV, it is essential that the learning model captures the contextual information about RV which is implicit in cardiac MRI scans i.e., its spatial closeness to the LV. Hence, in the supervised learning framework, we also utilized voxel-wise labels of the LV which were also provided in the training cohort.

**Table 1.** A summary of Multi-Disease, Multi-View, and Multi-Center RV segmentation in cardiac MRI (M&Ms-2021) dataset. Acronyms are: end-diastolic phase (ED), end-systolic phase (ES), short-axis view (SA), and long-axis view (LA).

| Dataset | Multi-Disease, Multi-View, and Multi-Center Cardiac MRI scans | | |
|---|---|---|---|
| **Centers** | Three clinical centers from Spain | | |
| **Scanners** | Siemens, Philips, and General Electric | | |
| **Cohorts (subjects)** | Training (160), Validation (40), and Challenge (160) | | |
| **Imaging data** | 3D, SA, Cardiac MRI scans of the ED and ES cardiac phases | | |
| | 2D, LA, Cardiac MRI scans of the ED and ES cardiac phases | | |
| **Segmentation data** | 3D segmentation maps on ED and ES cardiac phases in SA view | | |
| *Only available for training cohort* | 2D segmentation on ED and ES cardiac phases in LA view | | |
| **Segmentation Labels** | LV blood pool, RV blood pool, and LV Myocardium | | |
| **Pathology** | **Training Cohort** | **Validation Cohort** | **Challenge Cohort** |
| Normal subjects | 40 | 5 | 30 |
| Dilated Left Ventricle | 30 | 5 | 25 |
| Hypertrophic Cardiomyopathy | 30 | 5 | 25 |
| Congenital Arrhythmogenesis | 20 | 5 | 10 |
| Tetralogy of Fallot | 20 | 5 | 10 |
| Interatrial Comunication | 20 | 5 | 10 |
| Dilated Right Ventricle [*] | – | 5 | 25 |
| Tricuspidal Regurgitation [*] | – | 5 | 25 |
| [*] Tricuspidal Regurgitation (30 subjects) and Congenital Arrhythmogenesis (30 subjects) are only present in the validation and testing cohorts to evaluate generalization to unseen pathologies. | | | |

We make one interesting modification. We relabel the manual segmentation maps by merging the LV blood pool and LV myocardium, now assigned label=**1**, and assigned label=**2** to the RV. Since the focus of M&Ms-2021 challenge is accurate segmentation of RV, this modification utilized the prior information of spatial closeness of LV and RV without the need of learning to subdivide LV into LV blood pool and LV myocardium.

### 2.3 Multi-view SA-LA model

The proposed multi-view SA-LA model is based on a 2D U-Net architecture. It is formed with two U-Net architectures with asymmetrically large encoding-decoding pathways. One encoder-decoder pair segments RV on the short-axis view and the other pair on the long-axis view. The two U-Net architectures communicate encoded feature maps, extracted from matched SA and LA cardiac MR images, at the root of the encoders i.e., the highest down sampling level in the encoder branches. The synergistic information collected at the root of the encoder facilitates more accurate segmentation of RV on the SA and LA views. **Figure 1** shows a schematic of the proposed multi-view SA-LA architecture.

Each 2D U-Net comprised an encoding and a decoding path, with five down/up-sampling levels each. Feature maps at each level were processed by two $3 \times 3$ convolution layers, each followed by Batch Normalization and ReLU activation. Max-pooling and Transposed Convolutional layers served as down-sampling and up-sampling layers, respectively. Feature maps from the encoder of each U-Net were forwarded to the corresponding decoder, except for the highest down-sampling level (level 5) where each decoder received a concatenation of features maps from both encoders. To improve gradient propagation and feature learning in earlier layers, we employed deep supervision on the last three layers of the decoding pathways.

### 2.4 Training

The matched SA-LA dataset, elaborated in Section 2.2, formed the training cohort in our experiments. The proposed SA-LA model was trained with five-fold cross-validation to reduce data selection biases. Random batches of 2D slices were sampled from the matched SA-LA dataset followed by data augmentation procedure comprising the following operations: random horizontal and vertical flipping, random rotations, and random zooming. The SA-LA model was initialized with He-normal weight and trained for 150 epochs with an unweighted sum of Soft-Dice Loss and Cross-Entropy, Adam optimizer, and a constant learning rate of $1 \times 10^{-4}$.

### 2.5 Inference

Inference on the validation cohort was performed using the models learned with five-fold cross-validation on the training cohort. For each training fold, the optimal model for inference was selected based on the minimum validation loss[*]. For each subject in the validation cohort, the optimal model from each training fold was used to generate multi-class segmentation probability maps for LV and RV blood pool. The obtained probability maps from each fold were averaged to generate the final segmentation map for the SA and LA views. In the post-processing step, cluster thresholding was applied to remove isolated over-segmentations.

---

[*] Loss on local validation dataset (20% of training dataset)

Inference over the challenge cohort was performed by the challenge organizers. Participants, in the M&Ms-2021 challenge, were asked to provide the inference code and trained models in a docker environment which were ultimately used to generate segmentation maps on the challenge cohort.

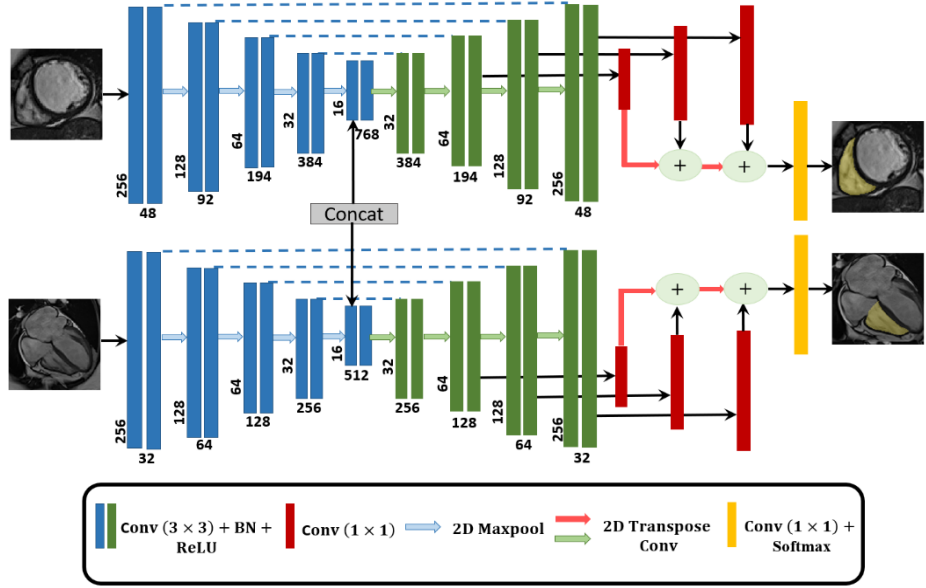

**Figure 1:** A schematic of the proposed multi-view SA-LA model.

### 2.6 Evaluation Metrics

Segmentation performance was quantified based on the Dice Similarity Coefficient (DSC in %) and Hausdorff Distance (HD-95 in mm). We computed DSC and HD for predicted segmentation maps in the ED and ES cardiac phases in SA and LA views. Firstly, for SA and LA views, DSC and HD scores were averaged across the ED and ES cardiac phases as follows:

$$\text{DSC}_{\text{SA}} = \frac{1}{2}\left(\text{DSC}_{\text{SA}}^{\text{ED}} + \text{DSC}_{\text{SA}}^{\text{ES}}\right) \text{ and } \text{HD}_{\text{SA}} = \frac{1}{2}\left(\text{HD}_{\text{SA}}^{\text{ED}} + \text{HD}_{\text{SA}}^{\text{ES}}\right)$$

$$\text{DSC}_{\text{LA}} = \frac{1}{2}\left(\text{DSC}_{\text{LA}}^{\text{ED}} + \text{DSC}_{\text{LA}}^{\text{ES}}\right) \text{ and } \text{HD}_{\text{LA}} = \frac{1}{2}\left(\text{HD}_{\text{LA}}^{\text{ED}} + \text{HD}_{\text{LA}}^{\text{ES}}\right) \quad (1)$$

Secondly, a unified score was computed by using the following formula provided by the M&Ms-2021 challenge:

$$\text{score} = \frac{0.75\big(\text{DSC}_{\text{SA}} + \widehat{\text{HD}}_{\text{SA}}\big) + 0.25\big(\text{DSC}_{\text{LA}} + \widehat{\text{HD}}_{\text{LA}}\big)}{2} \qquad (2)$$

where $\widehat{\text{HD}}_{\text{SA}}$ and $\widehat{\text{HD}}_{\text{LA}}$ are the normalized HD scores across participants in the challenge. This unified score was used to rank various methods proposed for segmentation of RV on cardiac MRI scans.

**Table 2.** Quantitative comparison of a non-holistic 2D U-Net model, multi-view SA-LA model without LV context, and (the proposed) multi-view SA-LA model (with LV context) on training, validation, and challenge cohorts. Score is computed using equation (2). Inferences are performed with an ensemble of five models learned with five-fold cross-validation on training cohort.

| Cohort (# subj) | Model | Parameters (millions) | DSC (%) | | | HD-95 (mm) | | | Score (%) |
|---|---|---|---|---|---|---|---|---|---|
| | | | Overall | SA | LA | Overall | SA | LA | |
| Training Cohort (160) | Non-holistic 2D U-Net model | ~39 | 90.14 | **89.99** | 90.59 | 5.84 | 6.69 | 3.28 | – |
| | Multi-view SA-LA model w/o LV context | ~27.2 | 89.62 | 89.42 | 90.25 | 4.96 | 5.82 | 2.36 | – |
| | Proposed Multi-view SA-LA model (with LV context) | ~27.2 | **90.60** | 89.59 | **93.61** | **2.89** | **3.16** | **2.07** | – |
| Validation Cohort (40) | Non-holistic 2D U-Net model | ~39 | 90.00 | 89.65 | **91.08** | 10.19 | 11.52 | 6.21 | 89.65 |
| | Multi-view SA-LA model w/o LV context | ~27.2 | 90.00 | 89.92 | 90.26 | 10.20 | 11.42 | 6.60 | 89.92 |
| | Proposed Multi-view SA-LA model (with LV context) | ~27.2 | **90.46** | **90.28** | 91.00 | **9.69** | **10.91** | **6.01** | **90.28** |
| Challenge Cohort (160) | Proposed Multi-view SA-LA model (with LV context) | ~27.2 | 90.66 | 91.00 | 89.63 | 10.40 | 11.16 | 8.13 | 91.00 |

## 3 Results

All experiments were performed on a system with 64GB RAM and an NVIDIA RTX 2080Ti 11GB GPU using open-source packages including Keras, Nibabel, Opencv, Numpy, Scikit-image, Pandas, and Matplotlib.

### 3.1 Quantitative Analysis

**Table 2** provides an extensive summary of quantitative results from various experiments conducted on the M&Ms-2021 dataset. We compared our proposed multi-view SA-LA model with a non-holistic 2D U-net model and a multi-view SA-LA model without LV context. The proposed multi-view SA-LA model had considerably fewer trainable parameters (~27 million) compared to a non-holistic 2D U-net model (~39 million) which trained two independent 2D U-net architectures for segmentation of RV in SA and LA views.

On the training and validation cohorts we found that the proposed multi-view SA-LA model outperformed other approaches in terms of DSC and HD metrics. More specifically, on the validation cohort (which included novel RV related pathologies), the proposed multi-view SA-LA model yielded the highest score (90.3%). Our experiments also revealed that adding LV contextual information boosted the segmentation performance on training and validation cohorts. On the validation cohort, we found that the multi-view SA-LA model (with LV context) increased the DSC by 0.46% and decreased the HD by 0.51mm.

### 3.2 Qualitative Analysis

**Figure 2** shows predicted segmentation maps of the RV for subject 153 in the training cohort. The non-holistic 2D U-Net model under-segments the basal, midventricular, and apical slices in the SA view. A multi-view SA-LA model without LV context substantially improved the segmentation of RV in the SA view. The proposed multi-view SA-LA model (with LV context) predicts segmentation maps closer to manual segmentations as quantified by DSC and HD metrics.

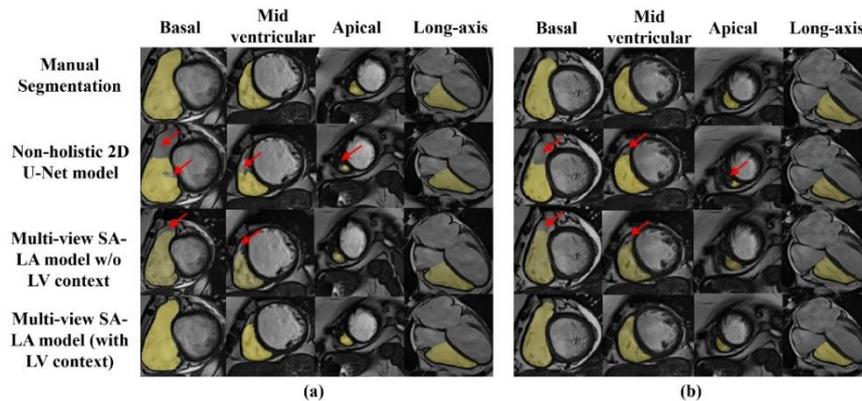

**Figure 2:** Predicted segmentation maps of the RV in the (a) end-diastolic and (b) end-systolic cardiac phases. The basal, midventricular, and apical slices of the short-axis (SA) view are shown in the first three columns of (a) and (b). A 2D slice in the long-axis (LA) view is shown in the fourth column of (a) and (b). Manual segmentation maps are shown in the first row for reference. Rows 2 to 4, in (a) and (b), show predicted segmentation maps obtained with a non-holistic 2D U-Net model, a multi-view SA-LA model without LV context, and a multi-view SA-LA model (with LV context). Red arrows show under-segmented regions in the predicted segmentation maps.

### 3.3 Generalizability Study on the Challenge Cohort

The proposed multi-view SA-LA model (with LV context) performed exceedingly well (score 91%) on the challenge cohort composed of cardiac imaging scans acquired at different centers with diverse cardiac pathologies. **Table 3** summarizes quantitative performance of the multi-view SA-LA model (with LV context) across eight cardiac pathologies in the challenge cohort. We find that, across pathologies, the DSC is consistently higher in the SA view ($\geq 90\%$) and the HD-95 is consistently lower in the LA view ($\leq 9$ mm). Our proposed segmentation approach also achieved superior segmentation performance on unseen RV related pathologies including Dilated Right Ventricle (DSC: SA 91.41, LA 89.63) and Tricuspidal Regurgitation (DSC: SA 91.40, LA 90.40) with low variance ($\sigma_{\text{DSC}}$: SA < 5%, LA< 6%).

## 4    Conclusion

In this study, we proposed a multi-view SA-LA model (with LV context) for simultaneous segmentation of RV on short-axis and long-axis views. SA-LA model has a multi-encoder, multi-decoder U-Net shaped structure where one encoder-decoder pair segments the RV blood on SA images and the other pair on LA images. We demonstrated, with extensive experiments on M&Ms-2021 dataset, that the proposed SA-LA model achieves state-of-the-art performance in the segmentation of RV on multi-view and multi-phase cardiac images. Moreover, SA-LA model (with LV context) shows high generalizability on novel RV related pathologies not represented in the training cohort.

**Table 3:** Quantitative performance of the proposed multi-view SA-LA model (with LV context) on eight different cardiac pathologies represented in the challenge cohort. 50 subjects with Dilated Right Ventricle and Tricuspidal Regurgitation help quantify generalizability of the segmentation algorithm as they are not represented in the training cohort. Acronyms are: DSC in short-axis view (DSC-SA), DSC in long-axis view (DSC-LA), HD-95 in short-axis view (HD-SA), and HD-95 in long-axis view (HD-LA).

| Pathology | No of Subjects | DSC-SA | DSC-LA | HD-SA | HD-LA |
|---|---|---|---|---|---|
| Normal subjects | 30 | $90.52 \pm 5.21$ | $91.40 \pm 5.72$ | $9.95 \pm 5.32$ | $6.22 \pm 5.05$ |
| Dilated Left Ventricle | 25 | $91.73 \pm 7.63$ | $86.80 \pm 17.12$ | $12.03 \pm 5.97$ | $7.10 \pm 5.60$ |
| Hypertrophic Cardiomyopathy | 25 | $89.99 \pm 8.39$ | $90.06 \pm 7.44$ | $11.80 \pm 7.59$ | $5.97 \pm 2.95$ |
| Congenital Arrhythmogenesis | 10 | $91.92 \pm 6.07$ | $90.81 \pm 7.22$ | $11.60 \pm 8.83$ | $7.15 \pm 5.98$ |
| Tetralogy of Fallot | 10 | $91.65 \pm 3.72$ | $90.74 \pm 3.30$ | $11.56 \pm 3.65$ | $7.81 \pm 3.65$ |
| Interatrial Comunication | 10 | $89.55 \pm 6.87$ | $86.17 \pm 17.34$ | $12.11 \pm 4.43$ | $9.44 \pm 9.55$ |
| **Dilated Right Ventricle** | 25 | $91.41 \pm 4.64$ | $89.63 \pm 5.31$ | $11.59 \pm 4.73$ | $8.89 \pm 5.59$ |
| **Tricuspidal Regurgitation** | 25 | $91.40 \pm 4.10$ | $90.40 \pm 4.52$ | $9.97 \pm 3.83$ | $6.80 \pm 4.31$ |